# Epitaxial Growth of Multiferroic YMnO$_3$ on GaN

*A. Posadas, J.-B. Yau, J. Han[*], C.H. Ahn*
Department of Applied Physics, Yale University, New Haven, Connecticut
[*]Department of Electrical Engineering, Yale University, New Haven, Connecticut

*S. Gariglio*
DPMC, University of Geneva, 24 Quai Ernest Ansermet, 1211 Geneva 4, Switzerland

*K. Johnston, J. B. Neaton[+], K.M. Rabe*
Department of Physics and Astronomy, Rutgers University, Piscataway, New Jersey
[+]The Molecular Foundry, Materials Sciences Division, Lawrence Berkeley National Laboratory, Berkeley, California 94720 and Department of Physics, University of California at Berkeley, Berkeley, California 94720

**In this work, we report on the epitaxial growth of multiferroic YMnO$_3$ on GaN. Both materials are hexagonal with a nominal lattice mismatch of 4%, yet x-ray diffraction reveals an unexpected 30° rotation between the unit cells of YMnO$_3$ and GaN that results in a much larger lattice mismatch (10%) compared to the unrotated case. Estimates based on first principles calculations show that the bonding energy gained from the rotated atomic arrangement compensates for the increase in strain energy due to the larger lattice mismatch. Understanding the energy competition between chemical bonding energy and strain energy provides insight into the heteroepitaxial growth mechanisms of complex oxide-semiconductor systems.**

The ubiquity of Si-based semiconductor devices today results in part from the ability to grow a high quality SiO$_2$ insulator to serve as the gate dielectric in field effect transistor applications. As various physical scaling limitations are approached, however, new high κ dielectric oxide materials are being explored to replace SiO$_2$. The concurrent development of similar, suitable oxide dielectrics for III-V and wide bandgap semiconductors is also being examined to enhance the functionality of these classes of semiconductors, as discussed by Doolittle and co-workers [1].

Here, we demonstrate the epitaxial growth of the complex oxide YMnO$_3$ on the wide bandgap semiconductor GaN. Both YMnO$_3$ and GaN have hexagonal symmetry and have a nominal lattice mismatch of about 4% (to half the lattice constant of YMnO$_3$). We observe an unexpected 30° rotation between the two unit cells, resulting in a much larger lattice mismatch of ~10 %. To understand these results, we have carried out first principles calculations of the elastic energy of highly strained YMnO$_3$. Because the relative stability of the first 1-2 layers determines the orientation of subsequent layers, we focus our calculations on the first unit cell at the interface. These calculations show that a large energy stabilization from bonding at the interface must overcome the elastic energy cost to result in the 30° rotation.

One potential advantage of using YMnO$_3$ for applications is its multiferroicity (i.e., simultaneous ferroelectricity and magnetism) [2]. Recently, thin films of YMnO$_3$ have been grown epitaxially on Si (111) [3-4]. These structures have been proposed for use in nonvolatile memory devices, taking advantage of the ferroelectric properties of YMnO$_3$ as well as its chemical stability at high temperatures. The mechanism of YMnO$_3$ epitaxy on GaN serves as a model for the development of epitaxial routes for other complex oxides grown on wide bandgap semiconductors, providing a test case for the development of multifunctional devices that incorporate diverse kinds of correlated behavior with semiconductors.

For these experiments, 2 µm thick, Ga terminated ((0001) face) epitaxial layers of GaN grown by metal-organic chemical vapor deposition (MOCVD) on C-plane sapphire wafers were used as substrates. X-ray diffraction (XRD) experiments reveal relaxed c-axis oriented growth of GaN, with a full-width-at-half-maximum (FWHM) of the rocking curve around the GaN 0002 peak of <0.2°. Prior to YMnO$_3$ growth, the GaN/sapphire was cleaned using standard electronic degreasing procedures.

YMnO$_3$ epitaxial films were grown by off-axis RF magnetron sputtering performed in a pure Ar atmosphere with a total pressure of 7.5x10$^{-2}$ torr and an RF power density of ~5 W/cm$^2$. The substrates were maintained at temperatures of 650-675°C during growth, as measured by an optical pyrometer, and were cooled to room temperature in vacuum after growth. Samples thicknesses ranged from 40-400 nm.

The as-grown films were characterized by atomic force microscopy (AFM), x-ray diffraction, and electrical and magnetic measurements. AFM scans of optimized films show a regular, continuous, smooth surface with a root-mean-square (RMS) roughness of 1-1.5 nm over an area of 10 µm x 10 µm. θ-2θ scans show peaks from the 000*l* reflections, indicating c-axis oriented growth (Fig. 1a). No impurity phases are detected. Rocking curves performed around the 0004 reflection yield a FWHM of 0.25°, similar to that of the GaN substrate, indicating high crystalline quality. To determine the in-plane orientation of the YMnO$_3$ film relative to GaN, off-axis $\phi$ scans were performed using the $10\bar{1}6$ reflection of YMnO$_3$ and the $10\bar{1}4$ reflection of GaN (Fig. 1b) [5]. The $\phi$ scans reveal the 6-fold symmetry of both the substrate and film, and an in-plane epitaxial relationship of $[11\bar{2}0]_{YMnO3}$ // $[10\bar{1}0]_{GaN}$ is found. From the diffraction scans, the YMnO$_3$ films are determined to be relaxed with lattice constants $a$ = 6.13 Å and $c$ = 11.38 Å. The striking feature about Figure 1b is the 30° rotation of the YMnO$_3$ lattice relative to the GaN lattice. This in-plane orientation produces a large lattice mismatch between YMnO$_3$ and GaN of ~10%, whereas for the unrotated case, the lattice mismatch is only ~4%. In order to understand the origin of the 30° rotation, we estimate the energies of formation of the rotated and aligned configurations, which is described in more detail below.

Ferroelectric characteristics of the YMnO$_3$ films were measured using standard electrical measurements. Gold contacts were deposited on samples that had been grown on Si-doped GaN/sapphire substrates ($10^{18}$ cm$^{-3}$ dopant concentration). The resulting ferroelectric hysteresis yields a coercive field of ~50 kV/cm and a remnant polarization of ~2 µC/cm$^2$ (Fig. 2). These values agree with recently reported values for YMnO$_3$ thin films on Si [3].

The magnetic properties were measured using a superconducting quantum interference device (SQUID) magnetometer. Under an applied field of 0.1 T, the in-plane magnetization of the sample was measured as a function of temperature in both field-cooled and zero field-cooled configurations. The magnetization shows Curie-Weiss behavior at high temperatures with a steep increase in the field-cooled measurement at ~45 K, indicating the Néel temperature of the YMnO$_3$ thin films.

We now turn to the discussion of the in-plane epitaxial relationship between YMnO$_3$ and GaN. To understand the observed orientation, we estimate the relative energies of formation of the unrotated interface and the 30° rotated interface. The most significant differences between the two interfaces involve the chemical bonding environment and the lattice mismatch. We find a competition between energy lowering from the saturation of Ga dangling bonds by O atoms and a concomitant increase in elastic strain energy required to fit the YMnO$_3$ lattice to the GaN lattice [6].

First, we consider the elastic strain energy. First-principles computations [7] were performed for strained bulk YMnO$_3$, with the measured strains imposed in the plane, relative to theoretical unstrained bulk YMnO$_3$. Other structural parameters were allowed to relax within the *P6$_3$mc* space group. For unstrained bulk YMnO$_3$ the calculated lattice parameters are $a_{theory}$ = 6.116 Å and $c_{theory}$ = 11.40 Å. For the unrotated case, the film strain [8], defined as $\dfrac{\left|\dfrac{a_{film}}{2} - a_{substrate}\right|}{\dfrac{a_{film}}{2}}$, is a 4.1% expansion. The resulting c lattice parameter is 1% larger than $c_{theory}$, and the elastic strain energy was calculated to be 0.158 eV per YMnO$_3$ formula unit. For the 30° rotation, the strain, defined as $\frac{\left|\frac{a_{film}}{\sqrt{3}} - a_{substrate}\right|}{\frac{a_{film}}{\sqrt{3}}}$, is much larger: a 9.9% compression. The resulting c lattice parameter is 7% larger than $c_{theory}$, and the elastic strain energy is 1.241 eV per YMnO$_3$ formula unit. It is necessary to calculate the strain energies explicitly as these large strains are outside the harmonic region. To illustrate this we also evaluated the strain energy using bulk elastic constants within the harmonic approximation. The computed elastic constants are $c_{11}$=317 GPa, $c_{12}$=204 GPa, $c_{13}$=87 GPa, and $c_{33}$=282 GPa (the bulk modulus is K=165 GPa, which is comparable to the measured bulk moduli of other common rare earth manganites [9]). Using the expression $W = \frac{1}{2}\varepsilon_i c_{ij} \varepsilon_j$, the strain energies are 0.34 eV and 1.96 eV for +4.1% and −9.9% mismatches, respectively, both significantly larger than the energies determined from the explicit calculations discussed above.

As a guide to the relative bonding energies at the interface between the GaN (0001) surface and the first unit cell layer of YMnO$_3$, we turn to reported first-principles calculations on the adsorption energies of oxygen atoms on GaN surfaces [10-12]. The energy gained from the bonding between the oxygen atoms and the Ga surface atoms depends on the degree of coverage of O atoms on the GaN surface and on the specific bonding sites at which the O atoms sit. There are four types of bonding sites on the GaN (0001) surface: the hcp site (on top of subsurface N), the fcc site (on top of the 3-fold cavity), the on-top site (on top of surface Ga), and the bridge site (in between two Ga surface atoms). The calculations show that at low coverages, the fcc site is the most energetically favorable oxygen bonding site. At high coverages, the adsorption energy is reduced by oxygen-oxygen repulsion [11], but this repulsion is expected to be strongly offset by the electrostatic interaction with the cations in the overlying YMnO$_3$ layers. For our estimates, we use the values of the adsorption energies per oxygen atom for each of the four bonding sites as tabulated in Ref. 12: 4.24 eV for hcp; 4.86 eV for on-top; 5.06 eV for bridge; and 6.38 eV for fcc.

In the unrotated configuration, the expected bonding is shown in Fig. 3 (bottom). In this case, the oxygen atoms cover 75% of the GaN (0001) surface. However, due to the arrangement of the oxygen atoms, the coverage is such that 1/3 of the oxygen atoms will sit in fcc sites, 1/3 in on-top sites, and 1/3 in hcp sites. This results in an average energy gain of 5.16 eV per YMnO$_3$ formula unit [13]. The expected bonding when the lattices are rotated by 30°, as observed experimentally, is shown in Fig. 3 (top). Here, the oxygen atoms cover 100% of the Ga (0001) surface, and the oxygen atoms are expected to sit on the minimum energy fcc sites. This bonding arrangement results in an energy gain of 6.38 eV per YMnO$_3$ formula unit [13]. Comparing the net energy gained from each interface configuration (chemical bonding energy minus strain energy), the 30° rotated case results in an energy gain of 0.15 eV per YMnO$_3$ formula unit (or 0.45 eV per YMnO$_3$ unit cell) over the aligned case, making the 30° rotated case more energetically favorable, in agreement with the experimental observations. These estimates indicate that energetic gains from the chemical bond formation can more than compensate for the cost of large epitaxial strain and stabilize the rotated configuration during the initial stages of growth.

In conclusion, we have grown high quality epitaxial thin films of multiferroic $YMnO_3$ on GaN/sapphire, with high quality structural properties having been obtained. We observe an epitaxial relationship of $[11\bar{2}0]_{YMnO3}$ // $[10\bar{1}0]_{GaN}$, which is determined to be the most stable configuration from calculations of the energy of formation of the interface. These calculations highlight the role of chemical bonding at the interface in determining the epitaxial relationship. The ability to grow epitaxial crystalline complex oxides such as $YMnO_3$ on GaN allows one to consider the development of multifunctional devices that couple correlated oxides with wide bandgap semiconductors. The availability of high quality thin films of multiferroic $YMnO_3$ also enables study on the origin of multiferroicity and the interaction between ferroelectricity and magnetism in this material.

We would like to thank Craig Fennie for valuable assistance. This work is supported by the Office of Naval Research under ONR grant N00014-02-1-0959 and Department of Energy grant DE-FG02-01ER45937.

**Figure Captions**

**Figure 1:** (a) X-ray diffraction θ–2θ scan of an YMnO$_3$ film showing c-axis oriented growth. The rocking curve around the 0004 peak has a full-width-at-half-maximum of 0.25°; (b) Off-axis ϕ scans of YMnO$_3$ ($10\bar{1}6$) and GaN ($10\bar{1}4$) reflections. The scans reveal a 30° in-plane rotational offset between the two unit cells.

**Figure 2:** Ferroelectric hysteresis loop of YMnO$_3$ with a coercive field of 50 kV/cm and a remnant polarization of 2 μC/cm$^2$.

**Figure 3:** Schematic of the in-plane orientational relationship between the GaN substrate and the YMnO$_3$ film for aligned unit cells with 4.1% tensile strain (bottom) and unit cells with a 30° offset with 9.9% compressive strain (top). The filled black circles are Ga surface atoms; the unfilled circles are N subsurface atoms of the substrate; and the filled blue circles are the atoms of the apical oxygen layer of YMnO$_3$. The black lines represent GaN unit cells, and the gray lines represent YMnO$_3$ unit cells.

# References

1. "Challenges and potential payoff for crystalline oxides in wide bandgap semiconductor technology," W.A. Doolittle, G. Namkoong, A.G. Carver, A.S. Brown, *Solid State Elec.* **47**, 2143 (2003).

2. "Why are there so few magnetic ferroelectrics?," N.A. Hill, *J. Phys. Chem. B* **104**, 6694-6709 (2000).

3. "Ferroelectric properties of YMnO$_3$ epitaxial films for ferroelectric-gate field-effect transistors," D. Ito, N. Fujimura, T. Yoshimura, T. Ito, *J. Appl. Phys.* **93**, 5563-5567 (2003).

4. "Ferroelectricity of YMnO$_3$ thin films on Pt(111)/Al$_2$O$_3$(0001) and Pt(111)/Y$_2$O$_3$(111)/Si(111) structures grown by molecular beam epitaxy," S. Imada, T. Kuraoka, E. Tokumitsu, H. Ishiwara, *Japn. J. Appl. Phys.* **40**, 666-671 (2001).

5. The $\phi$ scan was performed on a diffractometer without sample tilt perpendicular to the beam ($\chi$ angle), resulting in unequal intensities for symmetry related peaks.

6. In our model, the GaN is Ga terminated, and the subsequent layers are an apical oxygen layer of YMnO$_3$ followed by either a MnO layer or an Y layer. Bonding is assumed to occur between the Ga surface atoms and the apical oxygen layer of YMnO$_3$.

The formation of the YMnO$_3$-GaN interface then results from a competition between elastic strain and Ga-O bond formation in the first deposited layer.

7. First principles simulations were performed using the Vienna Ab-initio Simulation Package (VASP), which implements density functional theory and the LSDA+U method. The values U=8 and J=0.88 were taken from J. Medvedeva *et al.*, *J. Phys. Condens. Matter* **12**, 4947 (2000). Projector augmented-wave pseudopotentials were used, with the Y (4s, 4p, 5s, 4d), Mn (3p, 4s, 3d) and O (2s, 2p) treated as valence states. A plane wave energy cutoff of 600 eV and a 4x4x2 Γ-centered k-point sampling of the Brilllouin zone were used. Further details can be found in C.J. Fennie and K.M. Rabe, to be published.

8. Lattice mismatch is normally defined with respect to the substrate lattice constant. However, in calculations involving strain in the YMnO$_3$ film, the strain was defined with respect to the experimentally observed in-plane spacing of YMnO$_3$.

9. Available experimental single crystal elastic constant data for rare earth manganites are mostly for the colossal magnetoresistive materials (see for example K.G. Bogdanova *et al.*, *JETP Lett.* **80**, 308 (2004), C.H. Yang *et al.*, *J. Phys. Soc. Jpn.* **73**, 3051 (2004), V. Rajendran *et al., Phys. Status Solidi A* **195**, 350 (2003)). One published experimental measurement for bulk polycrystalline hexagonal manganites (M.C. Sekhar *et al.*, *Mod. Phys. Lett. B* **17**, 1119 (2003)) was found, with measured elastic moduli about eight times smaller than what we calculate here and what others have measured for single crystal colossal magnetoresistive materials.

10. "A theoretical study of O chemisorption on GaN (0001)/(0001) surfaces," J. Elsner, R. Gutierrez, B. Hourahine, R. Jones, M. Haugk, Th. Frauenheim, *Solid State Comm.* **108**, 953-958 (1998).

11. "The adsorption of oxygen at GaN surfaces," T.K. Zywietz, J. Neugebauer, M. Scheffler, *Appl. Phys. Lett.* **74**, 1695-1697 (1999).

12. "Theoretical study of $O_2$ adsorption on GaN surfaces," L. Nai-Xia, X. Yi-Jun, C. Wen-Kai, Z. Yong-Fan, *J. Molec. Struct. (Theochem)* **668**, 51-55 (2004).

13. There are 4 GaN unit cells per $YMnO_3$ unit cell in the aligned configuration, while there are 3 GaN unit cells per $YMnO_3$ unit cell in the 30° rotated configuration. There are 3 $YMnO_3$ formula units bonded to GaN per $YMnO_3$ unit cell.

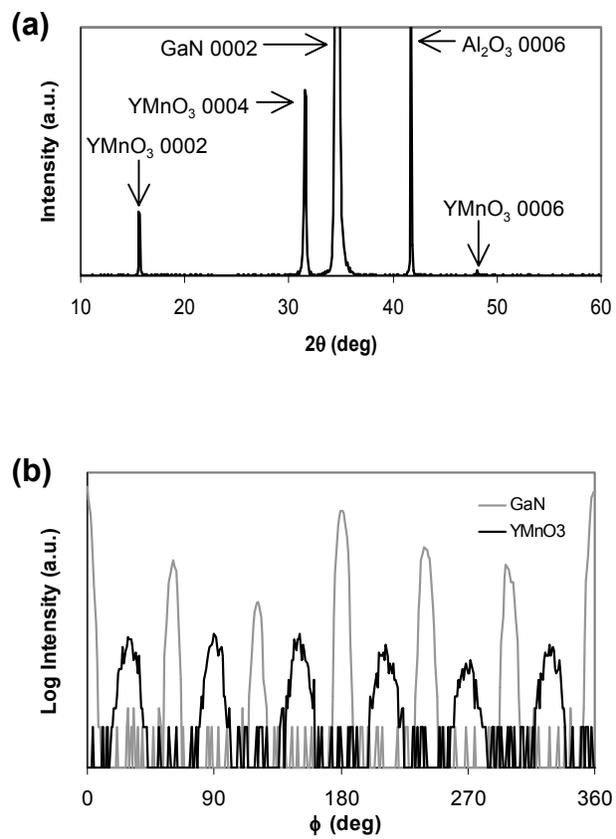

Figure 1

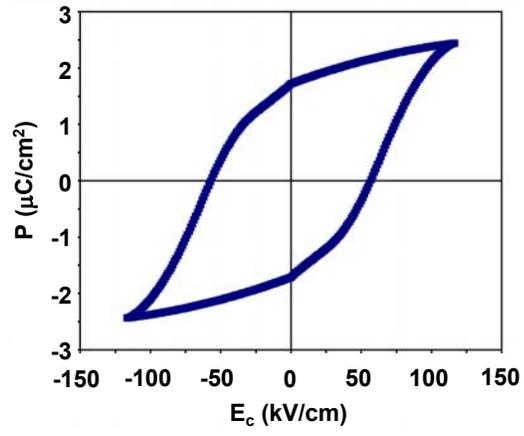

Figure 2

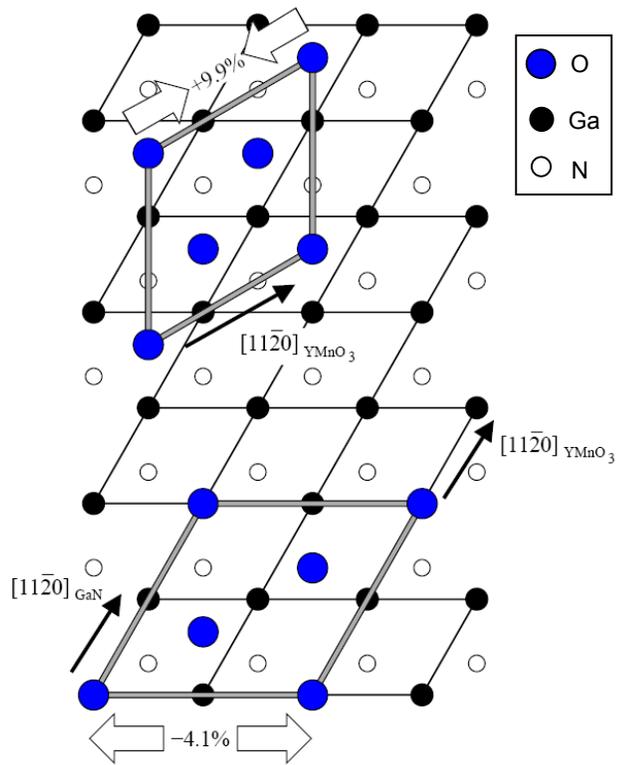

Figure 3